\documentclass[graybox]{svmult}

\usepackage{mathptmx}       
\usepackage{helvet}         
\usepackage{courier}        
\usepackage{type1cm}        
\usepackage{makeidx}         
\usepackage{graphicx}        
\usepackage{twocolumn}        
\usepackage[bottom]{footmisc}

\makeindex             

\begin{document}

\title*{Searching for galaxy clusters in the VST-KiDS Survey}
\titlerunning{Galaxy clusters in KiDS}

\author{M. Radovich$^1$, E. Puddu$^2$, F. Bellagamba$^3$, L. Moscardini$^3$, 
M. Roncarelli$^3$, F. Getman$^2$, A. Grado$^2$ and the KiDS collaboration\\
$^1$ INAF - OAPD\\
$^2$ INAF - OACN\\
$^3$ Dept. of Physics and Astronomy (DIFA), University of Bologna}

\authorrunning{M. Radovich et al.}

%
%
\maketitle

\abstract*{We present the methods and first results of the search for galaxy 
clusters in the Kilo Degree Survey (KiDS). 
The adopted algorithm and the criterium for selecting 
the member galaxies are illustrated. Here we report the preliminary results obtained over a 
small area ($7$ sq. degrees), and the comparison of our cluster candidates with those found in 
the RedMapper and SZ Planck catalogues; the analysis to a larger area ($148$ sq. degrees) is currently  
in progress. By the KiDS cluster search, we expect  to increase the completeness 
of the clusters catalogue to $z = 0.6-0.7$ compared to RedMapper.}

\abstract{We present the methods and first results of the search for galaxy 
clusters in the Kilo Degree Survey (KiDS). 
The adopted algorithm and the criterium for selecting 
the member galaxies are illustrated. Here we report the preliminary results obtained over a 
small area ($7$ sq. degrees), and the comparison of our cluster candidates with those found in 
the RedMapper and SZ Planck catalogues; the analysis to a larger area ($148$ sq. degrees) is currently  
in progress. By the KiDS cluster search, we expect to increase the completeness 
of the clusters catalogue to $z = 0.6-0.7$ compared to RedMapper.}

\section{Introduction}
\label{sec:1}

Clusters of galaxies are described as the most massive gravitationally bound quasi-equilibrium structures in 
the Universe. They represent a powerful tool for cosmology (\cite{Allen2011}): a large and complete catalogue 
of galaxy clusters spanning a broad range of redshifts 
would be important to understand the history of cosmic structure formation, and to provide accurate 
measurements of cosmological parameters, in particular $\sigma_8$ and $\Omega_M$. 
In this perspective, it becomes crucial the role of extragalactic surveys that can supply 
a statistically significant sample for the detection of clusters.
Among the already available surveys of different sizes and depths, one of the landmarks is the Sloan Digital
Sky Survey (SDSS; \cite{York2000}), which allows to probe the low-redshift Universe over an area of $20,000$ sq. 
degrees up to $z\sim 0.45$.   
The Kilo Degree Survey (KiDS; http://kids.strw.leidenuniv.nl/) and the Dark Energy Survey (DES; 
http://www.darkenergysurvey.org/) will allow to obtain samples of clusters spanning 
a wider range  of redshift and mass, thanks to their superior depth and image quality.
When completed, KiDS will cover 1500 sq. degrees in the $ugri$ bands, with optimal seeing conditions in the $r-$band; 
DES will cover a larger area (5000 sq. degrees in $gri$), with an image quality between SDSS and KiDS.
Here we show the very first results obtained from the comparison of clusters detected in KiDS with the RedMapper 
catalogue based on the SDSS (\cite{Rykoff2014}), containing $\sim 25,000$ clusters with masses $>10^{14}M_{\odot}$ 
in a volume limited up to $z<0.35$.

\begin{figure}[htbp]
\sidecaption
\includegraphics[scale=.35]{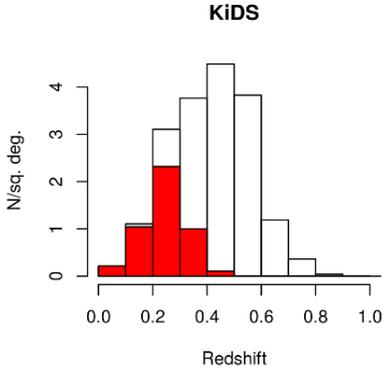}
\caption{Results from the Millennium Simulation derived with the KiDS photometric depths (black) 
and SDSS depths (red inset).
}
\label{fig:1}      
\end{figure}

\begin{figure}[b]
\sidecaption
\begin{tabular}{c}
\includegraphics[scale=.35]{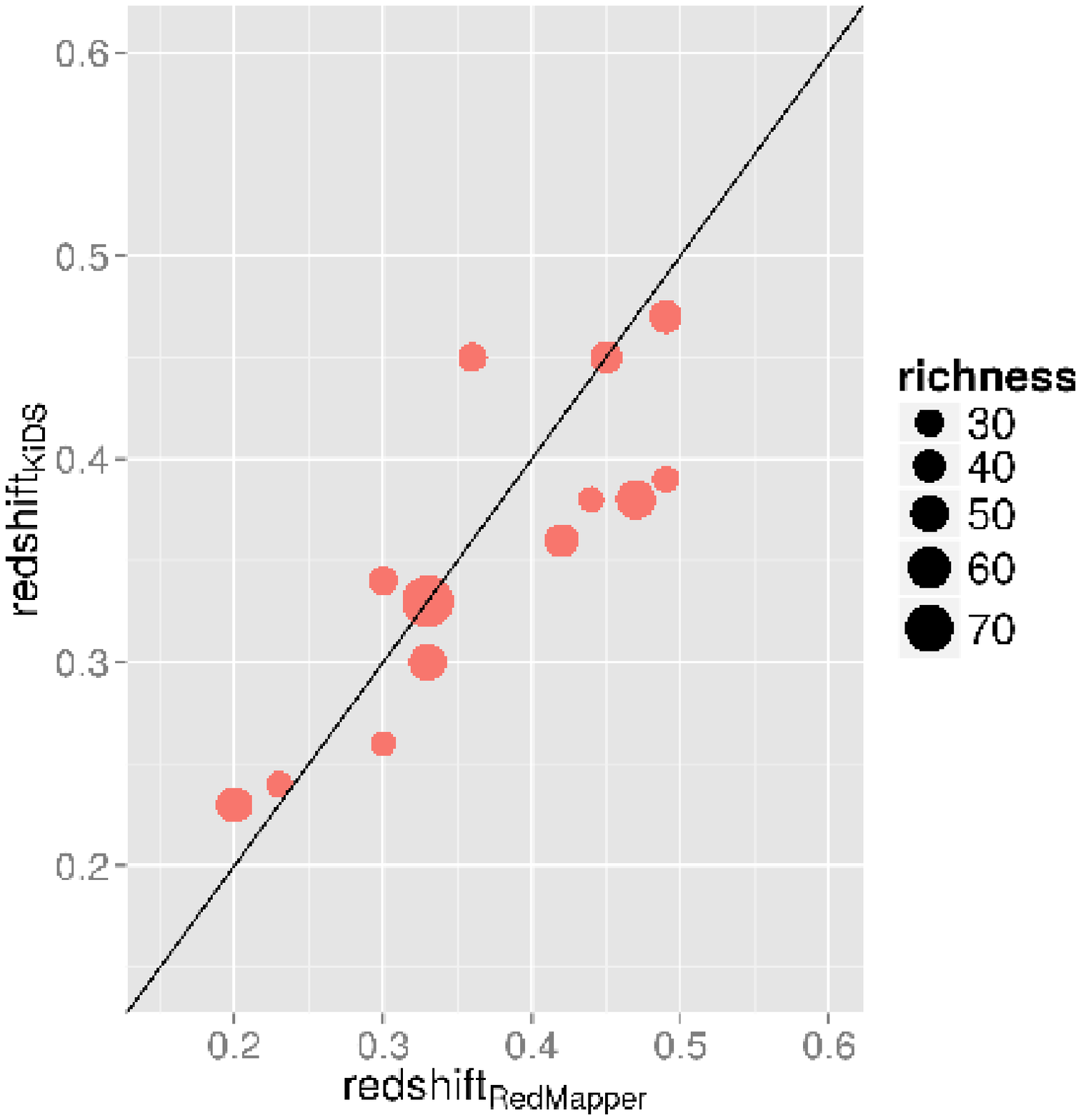}\\
\includegraphics[scale=.35]{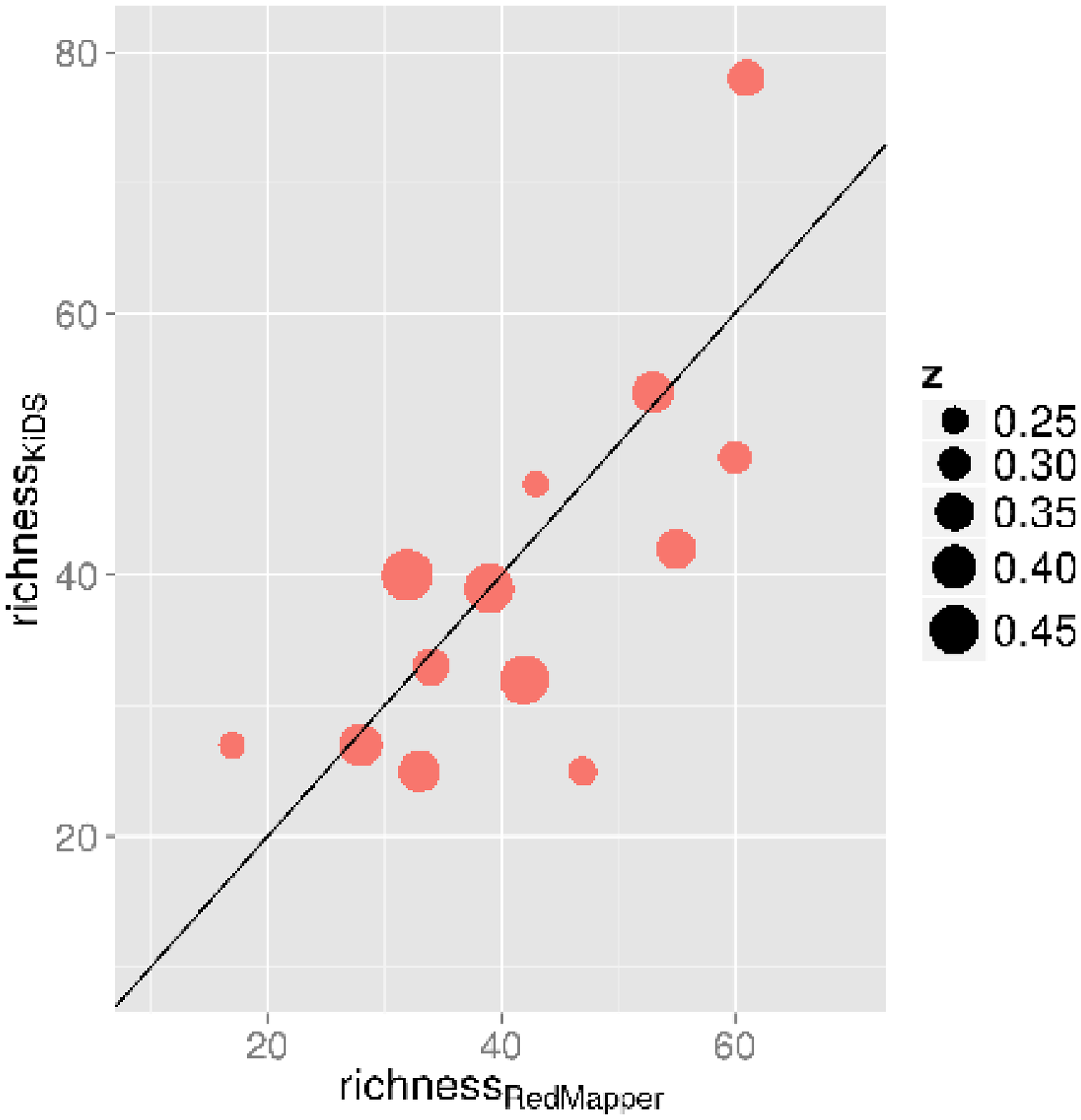}\\
\end{tabular}
\caption{Comparison of the photo-z and richness for the common KiDS/RedMapper clusters.}
\label{fig:2}       
\end{figure}

\section{The Data}
\label{sec:2}
The Kilo Degree Survey (\cite{DeJong2013}) is one of the ESO VST public surveys and is designed to observe 
an area of 1500 sq. degrees. in the {\it ugri} bands (with depths at 10 $\sigma$ respectively 
of 23, 24, 24 and 23 mags in the given bands). 
At present, a total 
of 148 sq. degrees, complete in {\it ugri}, have been published in the KiDS ESO Data Release DR2.
Data reduction was performed by the AstroWISE system (\cite{Verdoes2011}) within 
the KiDS consortium. 
Photometric redshifts were computed by the KiDS lensing team using the BPZ code (\cite{Benitez2010}) and 
benefit from the usage of the GAMA (\cite{Driver2011}) spectroscopy as a training set (for details
and photometric redshift accuracy, see \cite{Kuijken2015}).
An initial guess of the number of clusters vs. redshift expected in KiDS was derived using the 
mock catalogues by \cite{Henriques2012} from the Millennium Simulation (\cite{Springel2005}) 
assuming KiDS photometric depths. 
As it can be seen from Fig. \ref{fig:1}, KiDS should detect many more clusters at intermediate 
redshifts ($0.4 < z < 0.6$) with respect to SDSS (red inset). 
However, as it is described e.g. in \cite{Ascaso2014}, the semi-analytic galaxy formation models are not 
fully representing the photometric properties of galaxies in clusters; therefore, we are preparing more 
appropriate simulations built on KiDS data in order to estimate the completeness and purity of our cluster catalogue.

\begin{figure}[pht]
	\sidecaption
	\begin{tabular}{c}
		RMJ122120.3+002604.0\\
		\includegraphics[scale=.35]{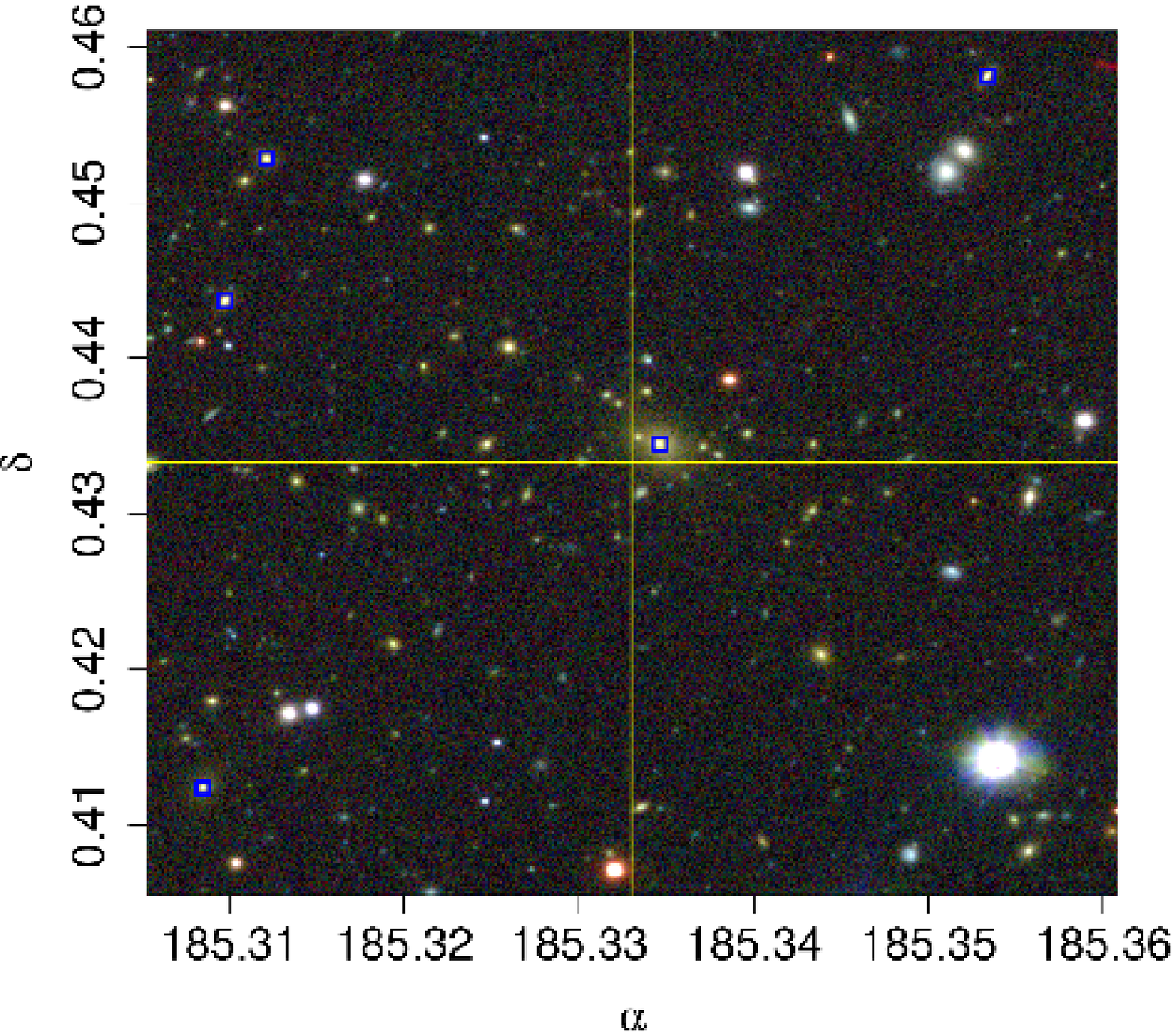}\\
		RMJ122428.6+005537.1 \\
		\includegraphics[scale=.35]{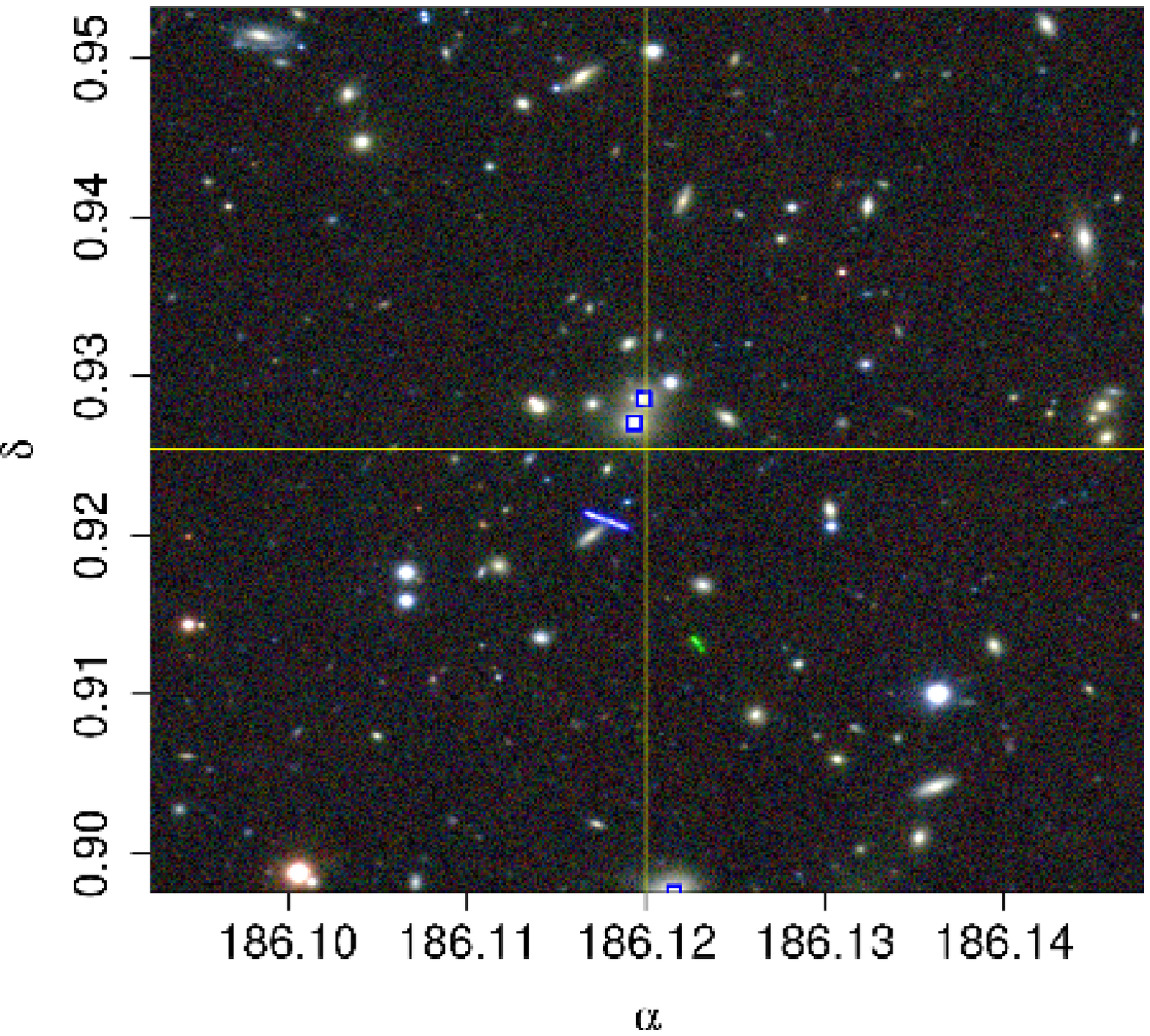} \\
		PSZ1 G011.93-63.54 \\
		\includegraphics[scale=.35]{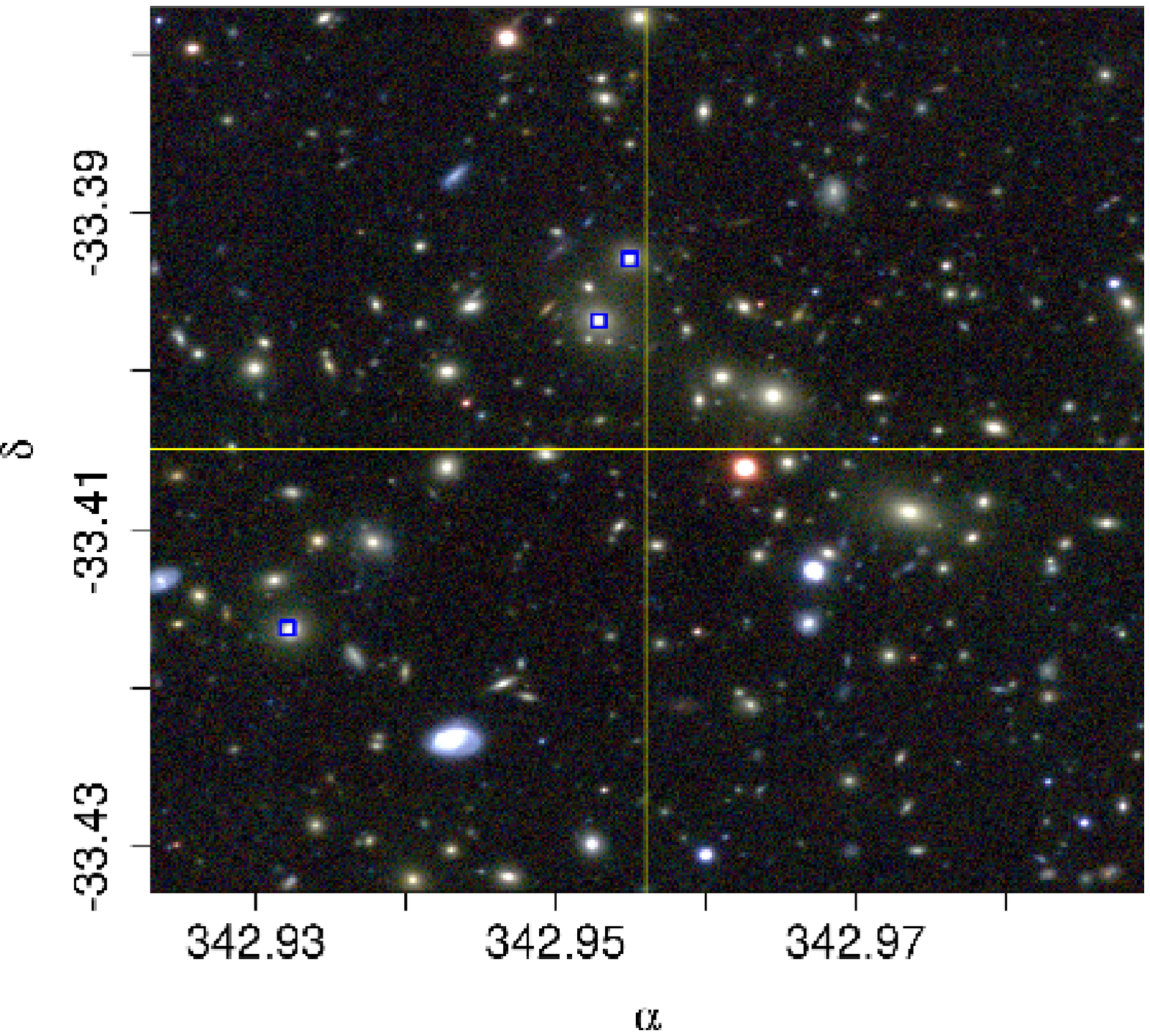}\\
	\end{tabular}
	\caption{{\it gri} KiDS image of three RedMapper/SZ Planck clusters. 
		The five brightest galaxies (if present in this central part of the field) are marked with blue squares.}
	\label{fig:3}       
\end{figure}

\begin{figure}[pht]
	\sidecaption
	\begin{tabular}{c}
		RMJ122120.3+002604.0  \\
		\includegraphics[scale=.2,bb=0 0 595 470]{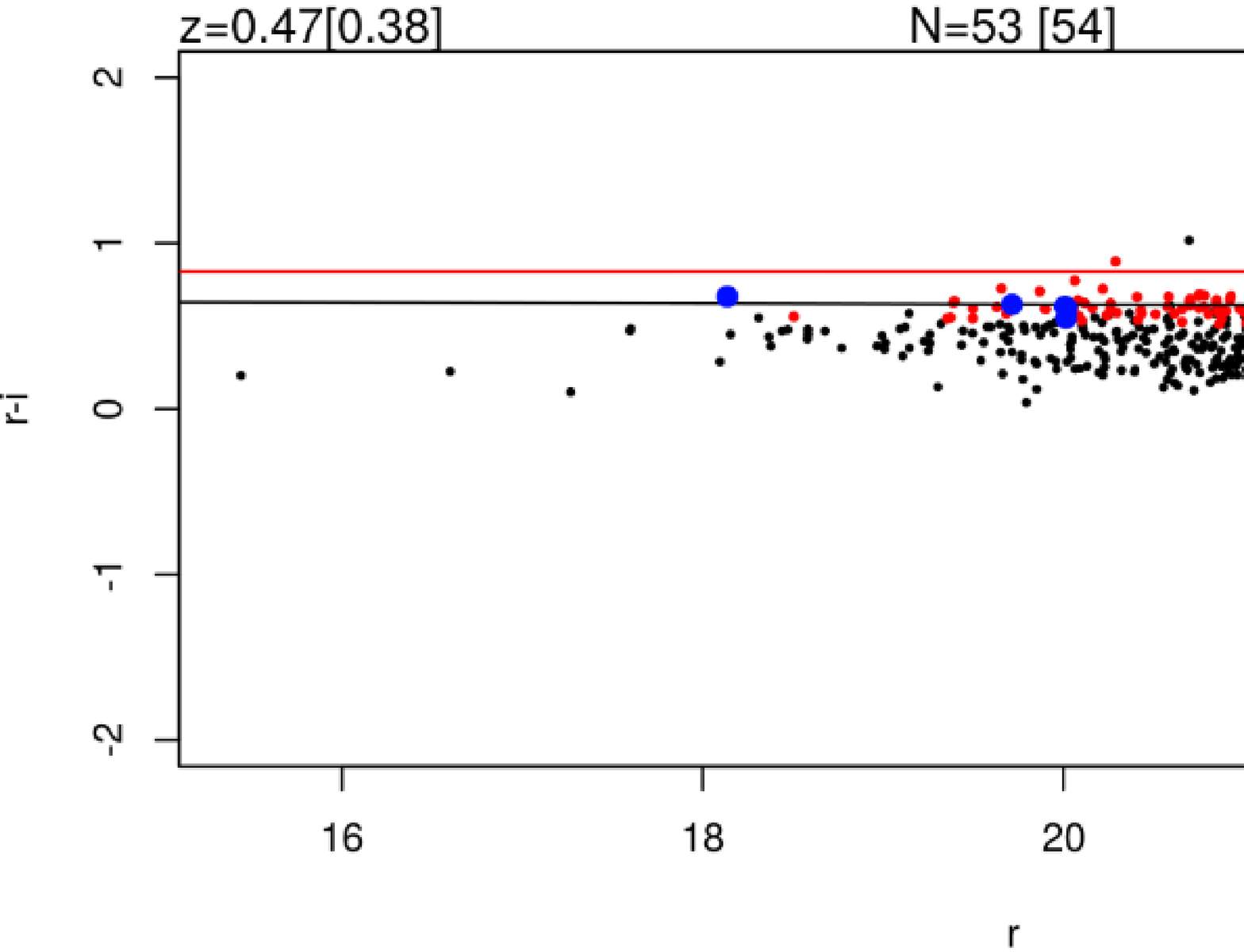}\\
		RMJ122428.6+005537.1 \\
		\includegraphics[scale=.2,bb=0 0 595 470]{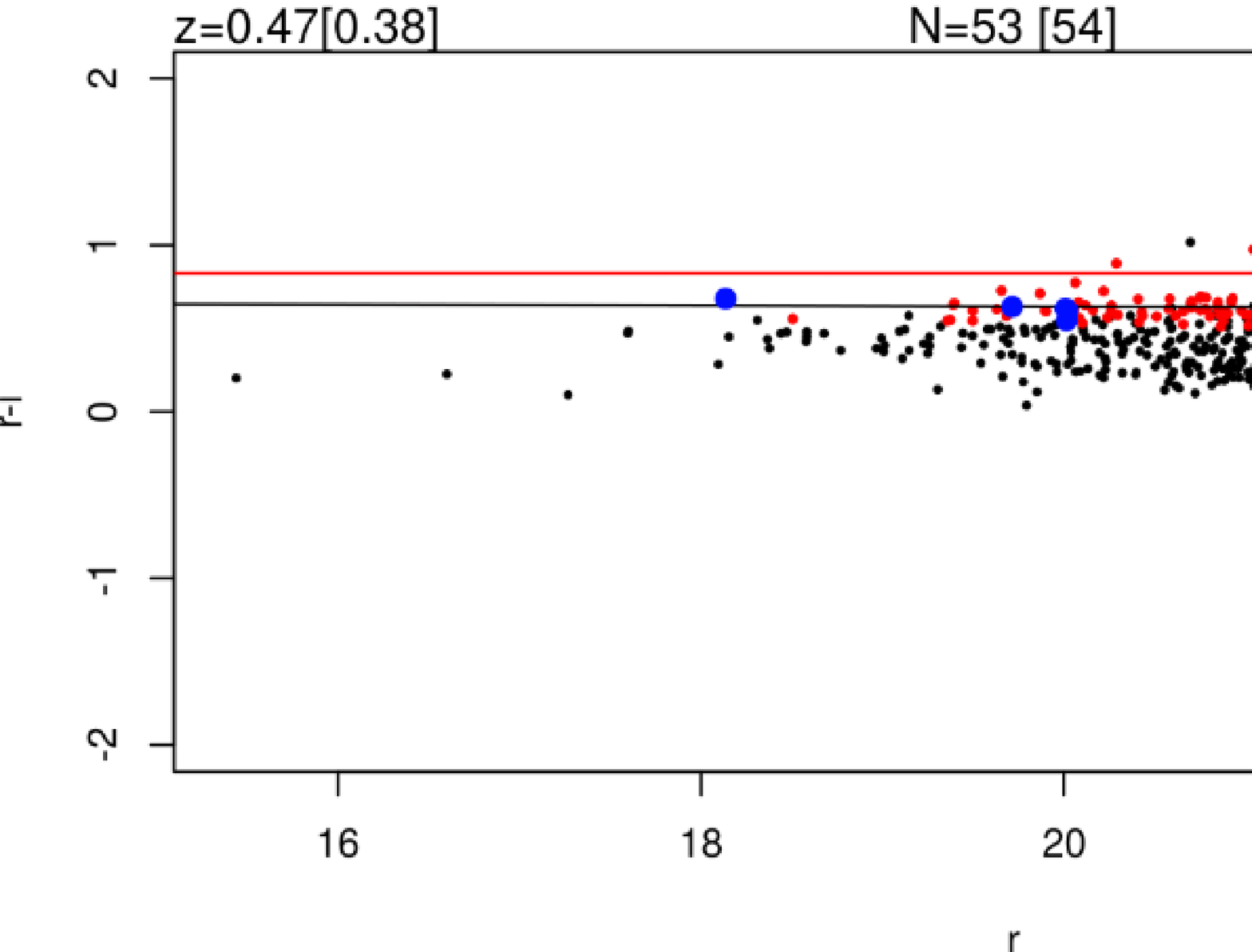}\\
		PSZ1 G011.93-63.54 \\
		\includegraphics[scale=.2,bb=0 0 595 470]{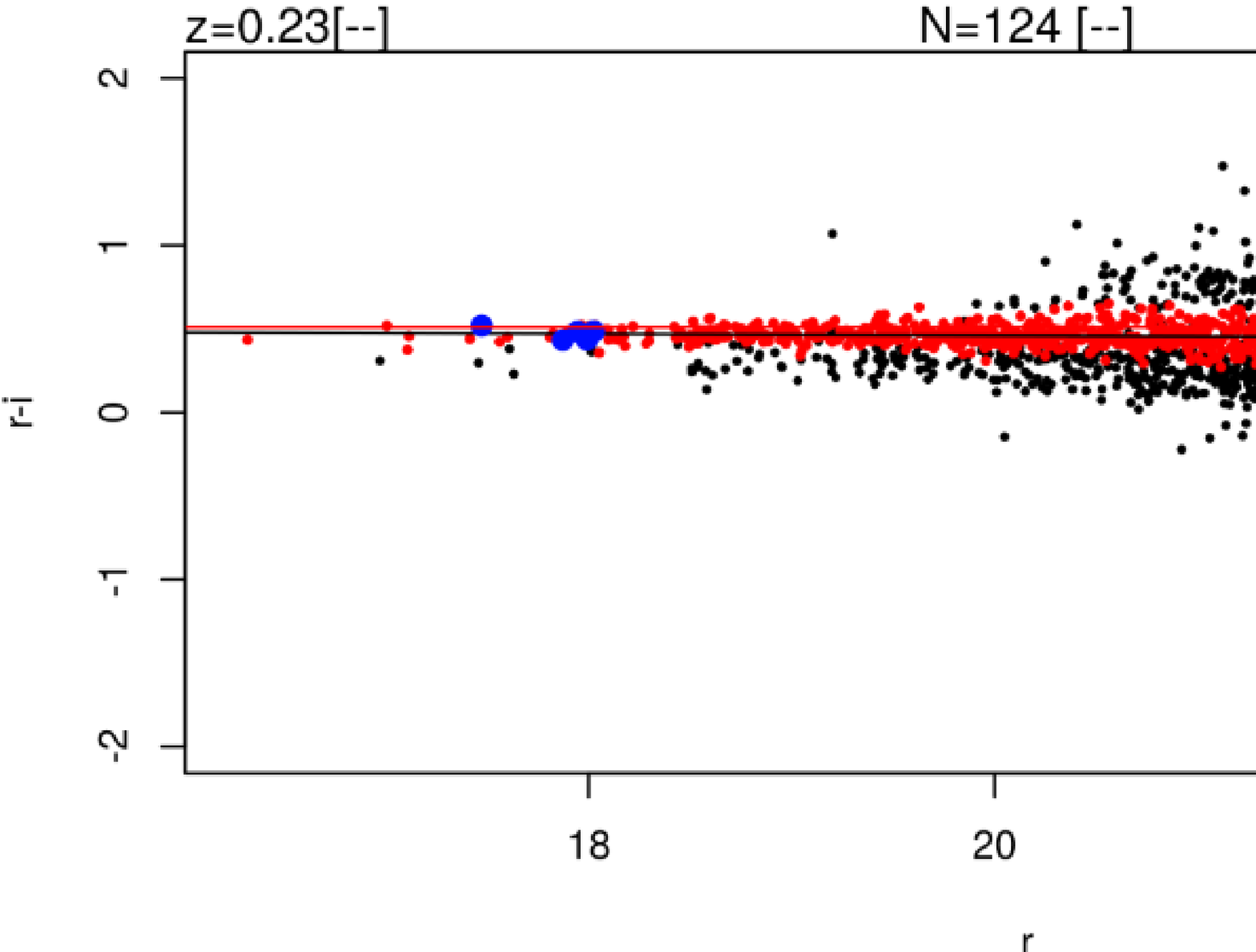}\\
	\end{tabular}
	\caption{r-i vs. r plot from KiDS data, for the clusters in Fig. \ref{fig:2}, of field (black) 
		and  Elliptical cluster member (red) galaxies. The five brightest galaxies are marked in blue 
		as in Fig. \ref{fig:2}. Redshift estimate, richness and signal to noise for the detection are reported
		on each panel; the RM values are also displayed in square brackets.}
	\label{fig:4}       
\end{figure}

\section{The Cluster Search}
\label{sec:3}

Cluster detection is performed with the Optimal Filtering technique, described in \cite{Bellagamba2011}. 
This method aims to maximise the signal-to-noise ratio of cluster detection by 
convolving the observed data with a suitable filter, which is proportional to the expected galaxy 
distribution in the clusters over the random fluctuations of the field distribution. 
This approach is very flexible, as it adapts to the properties of the different galaxy populations 
(cluster and field) in the available dataset, without selecting a priori a single signature of clusters.
In particular, in the analysis of the KiDS data, for each galaxy we use the position in the sky, the 
magnitude in the {\it r} band, and the full redshift probability distribution. For a set of redshift values, 
the cluster model is built from the mean properties of clusters in the MaxBCG sample derived from SDSS 
observations, appropriately modified to account for the evolution of the cluster galaxy population. 
The background is instead estimated from the mean density of galaxies in the field as a function of 
magnitude. The filter is then highlighting positions in the space where there are very luminous galaxies, 
which are more common in the clusters than in the field, and where the density is significantly higher 
than the mean one.

\section{Membership, BCGs and Richness}
\label{sec:4}

The algorithm provides for each cluster the center, the estimated redshift ($z_{CL}$) and 
the signal to noise ratio of the detection. 
In order to have a direct comparison with the RedMapper cluster catalogue 
 (which is based on the selection of Early-Type galaxies),
we need to define a similar galaxy membership criterium and apply it to our detections: to this purpose 
we rerun the photo-z code (BPZ) for all galaxies within $2 Mpc$ from the center with the redshift fixed 
to $z_{CL}$ and selected galaxies with a best-fit template consistent with Early-Type galaxies. 
Richness was, then, computed as the number of red galaxies placed within 1 Mpc from the center and 
with $L > 0.2L_*$ (see Fig. \ref{fig:3}, red dots).
We want to stress that these preliminary results depend on this specific data-set comprising a small
area; the procedure will be tuned as extensive simulations and more significant area data-set will be available.  
We initially run the cluster search on $7$ KiDS fields (a total area of $7$ sq. degrees) to check the 
performances of the cluster search algorithm and to validate it. 
There are 74 cluster candidates detected by our algorithm in this area. 
For comparison, in the same field there are 17 candidate clusters in the RedMapper catalogue, of which 13 
($76\%$) are common detections.
Furthermore, we find all the 3 ACOs (\cite{ACO89}) and the 2 XMMs (\cite{XMM2012}) clusters and
recover 3 of the 5 Planck clusters (\cite{Planck2014}).
Missing detections are located close to the masked regions of the fields, 
corresponding to the presence of bright foreground stars; clusters detected in KiDS and not in RedMapper are all at 
redshift $\geq 0.35-0.4$, at the detection limit of RedMapper.
In Fig. \ref{fig:2} we show the comparison of redshift and richness estimates for the KiDS 
candidate clusters with galaxy clusters found in the RedMapper survey.
In Fig.  \ref{fig:3}, we show the {\it gri} KiDS image of a
$3' \times 3'$ region around the cluster center for two RedMapper clusters and one 
SZ Planck cluster,  recovered from our algorithm on KiDS data. Fig.  \ref{fig:4}  displays  the related 
color-magnitude diagrams. 
Based on these preliminary results, an extension of this 
analysis to the 148 sq. degrees of the KiDS DR2 is currently in progress.


%

%
%
%

\end{document}